\documentclass[
aps
,prl
,preprint
,showpacs
]{revtex4-1}
\usepackage{epsfig}
\usepackage{bm}
\usepackage{color}
\usepackage{ulem}
\newcommand{\be}{\begin{eqnarray}}
\newcommand{\ee}{\end{eqnarray}}
\newcommand{\ba}{\begin{array}}
\newcommand{\ea}{\end{array}}
\newcommand{\bmat}{\left(\begin{array}}
\newcommand{\emat}{\end{array}\right)}
\newcommand{\no}{\nonumber}
\newcommand{\Tr}{\mbox{Tr}\,}

\begin{document}
\title{An exact solution of the partition function for mean-field quantum spin systems without the static approximation}
\author{Manaka Okuyama$^1$}
\author{Masayuki Ohzeki$^2$}%
\affiliation{%
$^1$Department of Physics, Tokyo Institute of Technology, Oh-okayama, Meguro-ku, Tokyo 152-8551, Japan
}
\affiliation{%
$^2$Graduate School of Information Sciences, Tohoku University, Sendai 980-8579, Japan
}

\date{\today}

\begin{abstract} 
Suzuki-Trotter decomposition is a well-known technique used to calculate the partition function of quantum spin systems, in which the imaginary-time dependence of the partition function occurs inevitably.
Since it is very difficult to explicitly treat the imaginary-time dependence of the partition function, we usually neglect the imaginary-time dynamical effect, which is called the static approximation.
Although the static approximation is the first approach, it is not even clear when the static approximation is justified for mean-field quantum spin systems, that is,  mean-field quantum spin systems have not been solved exactly so far.
In this study, we solve exactly the partition function for a particular class of mean-field quantum spin systems including randomness  without the static approximation.
The partition function can be regarded as a result of time evolution in the imaginary-time Schr\"odinger equation, and solving the exact solution of the  partition function is equivalent to solving the optimal control problem in the imaginary-time Schr\"odinger equation.
As the result, the solution of the optimal control problem coincides exactly with the static approximate solution of the partition function and, therefore, the static approximation is exact for the particular class of mean-field quantum spin systems including randomness in general.
Furthermore, we prove that the analysis of the previous study in quantum annealing is exact where the non-stoquastic interaction and the inhomogeneous transverse field  accelerate the computational time exponentially for mean-field quantum spin systems.
\end{abstract}

\maketitle

%%%%%%%%%%%%%%%%%%%%%%%%%%%%%%%%%%%%%%%%%%%%%%%%%%%%%%%%%%%%%%%%%%%%%%%%%%%
{\it Introduction.}---
Solving the partition function is the most important problem of statistical mechanics.
Unfortunately, it is a very difficult problem to obtain exactly the partition function  in general, and there are extremely few models whose partition function can be obtained.
In classical mean-field systems \cite{NO}, it is easy to obtain exactly the free energy and the self-consistent equation in the thermodynamic limit  by the saddle point method.
Using the Suzuki-Trotter decomposition, similar analysis is also effective for quantum mean-field systems.
However, at this time, since the partition function contains the imaginary-time dependence, it is difficult to analyze exactly
and we usually employ the static approximation \cite{BM} which neglects the imaginary-time dependence.
The static approximation is an indispensable method  as the zero-th order approximation, and the partition function of mean-field quantum spin systems can not be attained without the static approximation so far.

The static approximation does not necessarily give a good approximate solution and its exactness has been investigated for individual models.
For example, it is known that the static approximation does not hold for the $p$-body spin glass model with the transverse-field \cite{YI,WW,GL,Goldschmidt,UBK,MH,ONS,Takahashi}.
On the other hand, although previous study \cite{CCIL,KRSZ,JKKMP,SN} implies that the static approximation is exact for the infinite-range ferromagnetic $p$-spin model, its proof is not given in general.
There is no general understanding as to when the static approximation is exact.

In this Letter, using the method of the optimal control problem \cite{AP,Kolosov}, we derive exactly the partition function for a particular class of mean-field quantum spin systems including randomness without the static approximation.
The optimal control problem is originally developed in the field of control engineering and is a theory of determining the time dependence of the coefficients of the differential equations so as to minimize (or maximize) a given cost function.
Using the fact that the imaginary-time dependence of the partition function can be regarded as the result of time evolution of the imaginary-time Schr\"odinger equation \cite{Koh, Koh2}, we map obtaing the partition function to solving the optimal control problem in the imaginary-time Schr\"odinger equation.
Although it is very difficult to analytically solve the optimal control problem in general, we can find the optimal solution at this time. 
As the result, the solution of the optimal control problem, that is, the exact solution of the partition function coincides exactly with the static approximate solution of the partition function.
The above result is applicable to the particular class of mean-field quantum spins systems including randomness and, thus, the static approximation is exact for their models in general.
Our result gives for the first time a systematic method to exactly analyze mean-field  quantum spin systems, and
it is an important step for exact analysis beyond the static approximation.

In addition, our result is also very important from the viewpoint of quantum annealing \cite{KN,FGGS} where the computational time is closely related to quantum phase transition and the performance evaluation is carried out by analyzing the phase diagram of  Hamiltonians \cite{SN,SN2,SYYN,MNAL,OYNR}.
Previous studies \cite{SN,SN2,SYYN} show that, based on the static approximation, non-stoquastic interaction and inhomogeneous transverse fields exponentially improve the computational time of quantum annealing for mean-field quantum spin systems.
Our result includes their mean-field models and, therefore, prove that their results \cite{SN,SN2,SYYN} are exact.

%%%%%%%%%%%%%%%%%%%%%%%%%%%%%%%%%%%%%%%%%%%%%%%%%%%%%%%%%
%%%%%%%%%%%%%%%%%%%%%%%%%%%%%%%%%%%%%%%%%%%%%%%%%%%%%%%%%%%%%%%%%%%%%%%%%%%
{\it Exact solution of infinite-range ferromagnetic $p$-spin model.}---
First, we consider the infinite-range ferromagnetic $p$-spin model with the transverse-field term as a simple case,
\be
\hat{H}= -  N \left(\frac{1}{N^{}} \sum_i \hat{\sigma}_i^z \right)^p - \Gamma\sum_i  \hat{\sigma}_i^x ,
\label{inhom}
\ee
Following the standard procedure \cite{SN}, we obtain the partition function as follows, 
\be
Z&=&\lim_{M\rightarrow\infty}\int \mathcal{D} m^z \left\{  \Tr\left( \prod_{t=1}^M e^{  \frac{\beta}{M}   p(m^z(t))^{p-1} \hat{\sigma}^z } e^{ \frac{\beta}{M}\Gamma \hat{\sigma}^x  } e^{ -\frac{\beta}{M} (p-1)(m^z(t))^p  } \right) \right\}^N . \label{partition}
\ee
Since the trace of the partition function contains the imaginary-time dependence, it is difficult to proceed with further calculation without some approximation.
When we use the static approximation which neglects all the $t$ dependence of the parameter, we can take trace in Eq. (\ref{partition}) using the inverse operation of the Trotter decomposition. 
Then, we obtain the pseudo free energy and the saddle point equation,
\be
f_{\rm SA} &=&(p-1)   (m^z)^p-\frac{1}{\beta} \log\left\{  2\cosh\left( \beta \sqrt{ p^2(m^z)^{2p-2} +\Gamma^2} \right) \right\} ,
\\
m^z &=&p(m^z)^{p-1} \frac{\tanh\left(\beta \sqrt{p^2(m^z)^{2p-2} +\Gamma^2} \right)}{\sqrt{p^2(m^z)^{2p-2} +\Gamma^2} } . \label{saddle}
\ee
However, there is no guarantee that the static approximation is exact, and the following relation generally holds between the exact solution and the static approximate solution in the thermodynamic limit,
\be
 f \le f_{\rm SA} .
\ee
Our aim is to attain  the exact partition function  in the thermodynamic limit.
Of course, it is difficult to deal with the trace of Eq. (\ref{partition}) directly. Then, we convert the trace of Eq. (\ref{partition}) into the imaginary-time Schr\"odinger equation \cite{Koh, Koh2}. 
We consider the following imaginary-time Schr\"odinger equation,
\be
\frac{d}{ds} | \psi(s)\rangle &=&     
\left(\begin{array}{ccccc}
      \hat{H}_{IM}(s) & 0 \\
      0 & \hat{H}_{IM}(s)\\
    \end{array}
 \right)|
 \psi(s)\rangle  ,  \label{HIM}
 \\
 \hat{H}_{IM}(s)&=& p(m^z(s))^{p-1} \hat{\sigma}^z  + \Gamma \hat{\sigma}^x - (p-1) (m^z(s))^p ,
 \\
  |\psi(s) \rangle &=&
  \left(
    \begin{array}{ccccc}
      x_1(s) &  x_2(s)&  x_3(s) & x_4(s) 
          \end{array}
  \right) ^{\mathrm{T}}  ,
   \\
    |\psi(0) \rangle &=&
  \left(
    \begin{array}{ccccc}
      1 &  0&  0 & 1 
          \end{array}
  \right) ^{\mathrm{T}}  ,
\ee
where $0\le s \le \beta$.
Then, we find that the trace of Eq. (\ref{partition}) is equivalent to $ x_1(\beta) +x_4(\beta) $,
\be
\Tr\left( \prod_{t=1}^M e^{  \frac{\beta}{M}   p(m^z(t))^{p-1} \hat{\sigma}^z } e^{ \frac{\beta}{M}\Gamma \hat{\sigma}^x  } e^{ -\frac{\beta}{M} (p-1)(m^z(t))^p  } \right)
&=& x_1(\beta) +x_4(\beta)  .
\ee
So far, the imaginary-time dependence of $m^z(s)$ is arbitrary, and the partition function is calculated by summing up arbitrary paths. However, in the thermodynamic limit, only the path with the largest value of $x_1(\beta) +x_4(\beta) $ contributes to the partition function.
Therefore, if we can find the imaginary-time dependence of $m^z(s)$ such that the value of $x_1(\beta) +x_4(\beta) $ is the largest in Eq. (\ref{HIM}), the partition function can be exactly obtained in the thermodynamic limit. 
The problem of finding the time dependence of the coefficients of the differential equation so as to maximize the given cost function is equivalent to the optimal control problem and, in the following, we will find the optimal imaginary-time dependence of $m^z(s)$ using the method of the optimal control problem.

%%%%%%%%%%%%%%%%%%%%%%%%%%%%%%%%%%%%%%%%%%%%%%%%%%%%%%%%%
From the result of the optimal control problem \cite{AP,Kolosov}, a necessary condition for ${m^z}^\ast(s)$ and $x^\ast(s)$ to be optimal is that  there exists a function $k^\ast(s)$ such that simultaneously satisfies the following conditions (see supplemental material for detail),
\be
\dot{x}_i^\ast(s)&=& \frac{\partial H_{\rm op}^\ast }{\partial k_i} ( x^\ast(s), k^\ast(s) , {m^z}^\ast(s) )  ,
\\
\dot{k}_i^\ast(s)&=& -\frac{\partial H_{\rm op}^\ast }{\partial x_i} ( x^\ast(s), k^\ast(s) , {m^z}^\ast(s) )  ,
\\
x_1(0) &=& x_4(0) =1 , \label{b1}
\\
x_2(0) &=& x_3(0) =0 ,
\\
k_1(\beta)&=&k_4(\beta)=-1 ,
\\
k_2(\beta)&=&k_3(\beta)=0 , \label{b2}
\\
H_{\rm op}^\ast( x^\ast(s), k^\ast(s) , {m^z}^\ast(s) ) 
&=& \min_{m^z}  H_{\rm op}( x^\ast(s),  k^\ast(s), m^z(s) )  ,
\ee
where  the classical Hamiltonian $H_{\rm op}$ is given by 
\be
H_{\rm op} &=&-(p-1) (m^z)^{p} \sum_{i=1}^{4} x_i k_i  +  p (m^z)^{p-1} \sum_{i=1}^{4} (-1)^{i-1}  x_{i} k_{i}
\no\\
&&+ \Gamma  (x_{1} k_{2} + x_{2} k_{1} +x_{3} k_{4} + x_{4} k_{3}) . \label{Hcl}
\ee
From Eq. (\ref{Hcl}), we find
\be
{m^z}^\ast(s)=0, \ \frac{\sum_{i=1}^{4} (-1)^{i-1}  x_{i}^\ast(s) k_{i}^\ast(s) }{ \sum_{i=1}^{4} x_i^\ast(s) k_i^\ast(s) } . \label{min}
\ee 
As we will see later, since the nontrivial solution ${m^z}^\ast(s)=\sum_{i=1}^{4} (-1)^{i-1}  x_{i}^\ast(s) k_{i}^\ast (s)/ \sum_{i=1}^{4} x_i^\ast (s)k_i^\ast(s) $ contains the trivial solution ${m^z}^\ast(s)=0$, we focus only on the nontrivial solution.
Then, although Eq. (\ref{min}) is the necessary condition for ${m^z}^\ast(s)$, there is only one condition in  Eq. (\ref{min}).
Thus, if we find one solution, it is just the optimal solution from uniqueness of the optimal control.
Equation (\ref{Hcl}) is reduced to 
\be
H_{\rm op}^\ast&=&\Gamma (x_1 k_2 +x_2 k_1 + x_3 k_4 +x_4 k_3) +  \left(\frac{\sum_{i=1}^{4} (-1)^{i-1}  x_{i} k_{i} }{ \sum_{i=1}^{4} x_i k_i }\right)^p  \sum_{i=1}^{4} x_i k_i  .
\ee
Obviously, it is difficult to find a general solution of the Hamilton equations of $H_{\rm op}^\ast$ except for  $p=1$.
However, we will see that the solution of $H_{\rm op}^\ast$ can be just obtained in the case where the boundary conditions are given by Eqs. (\ref{b1})-(\ref{b2}).

From the Hamilton equations of $H_{\rm op}^\ast$, the equation of motion of ${m^z}^\ast(s)$ is given by
\be
\frac{d}{ds} {m^z}^\ast(s)=-\frac{2\Gamma(x_1^\ast k_2^\ast -x_2^\ast k_1^\ast + x_3^\ast k_4^\ast - x_4^\ast k_3^\ast) }{x_1^\ast k_1^\ast +x_2^\ast k_2^\ast +x_3^\ast k_3^\ast +x_4^\ast k_4^\ast}  .
\ee
Here, we consider the Hamilton equations of $H_{\rm op}$ when $m^z$ is constant.
We immediately find the solutions of $x_i^C(s)$ and $k_i^C(s)$ as follows,
\be 
 \left(
    \begin{array}{c}
      x_1^C(s) \\
      x_2^C(s) \\    
      x_3^C(s) \\
      x_4^C(s) \\
          \end{array}
  \right) 
     &=&
     e^{- s(p-1) (m^z)^p}    \left(
    \begin{array}{c}
      \cosh \left(s \sqrt{ p^2(m^z)^{2p-2}   + \Gamma^2 } \right )     + \frac{p(m^z)^{p-1} \sinh \left(s \sqrt{ p^2(m^z)^{2p-2}   + \Gamma^2 } \right )}{\sqrt{ p^2(m^z)^{2p-2}   + \Gamma^2 } } \\
      \frac{\Gamma \sinh \left(s \sqrt{ p^2(m^z)^{2p-2}   + \Gamma^2 } \right )}{\sqrt{ p^2(m^z)^{2p-2}   + \Gamma^2 } } \\    
     \frac{\Gamma \sinh \left(s \sqrt{ p^2(m^z)^{2p-2}   + \Gamma^2 } \right )}{\sqrt{ p^2(m^z)^{2p-2}   + \Gamma^2 } } \\
      \cosh \left(s \sqrt{ p^2(m^z)^{2p-2}   + \Gamma^2 } \right )     -\frac{p(m^z)^{p-1} \sinh \left(s \sqrt{ p^2(m^z)^{2p-2}   + \Gamma^2 } \right )}{\sqrt{ p^2(m^z)^{2p-2}   + \Gamma^2 } }\\
          \end{array}
  \right) ,
\no\\
\\
  \left(
    \begin{array}{c}
      k_1^C(s) \\
      k_2^C(s) \\    
      k_3^C(s) \\
      k_4^C(s) \\
          \end{array}
  \right)
     &=&
   -\left(
    \begin{array}{c}
      x_1^C(\beta-s) \\
      x_2^C(\beta-s) \\    
      x_3^C(\beta-s) \\
      x_4^C(\beta-s) \\
          \end{array}
  \right)  .
  \ee
  Then, we find
  \be
x_1^C(s) k_2^C(s) -x_2^C(s) k_1^C(s) + x_3^C(s) k_4^C(s) - x_4^C(s) k_3^C(s)=0 .
  \ee
  Thus, when we put ${m^z}(s)$ as 
\be
{m^z}^C(s)= \frac{\sum_{i=1}^{4} (-1)^{i-1}  x_{i}^C(s) k_{i}^C(s) }{ \sum_{i=1}^{4} x_i^C(s) k_i^C(s) } , \label{min2}
\ee
in the Hamilton equations of $H_{\rm op}$, then we find that ${m^z}^C(s)$ is constant
  \be
  \frac{d}{ds} {m^z}^C(s)=0 ,
  \ee
  and the solution of $H_{\rm op}$ are also given by $x_i^C(s), k_i^C(s)$ and ${m^z}^C(s)$.
  In addition, we find that this solution is simultaneously the solution of $H_{\rm op}^\ast$ because Eq. (\ref{min}) is satisfied.
  Therefore,  from uniqueness of the optimal control, this is just the solution of the motion of $H_{\rm op}^\ast$, {\it i.e.}, the solution of the optimal control problem is given by
\be
x_i^\ast(s)  &=& x_i^C(s) ,
 \\
k_i^\ast(s)  &=& k_i^C(s)  ,
    \ee
under the condition of Eq. (\ref{min2}).
Furthermore, the condition (\ref{min2}) reproduces the self-consistent equation of the static approximation (\ref{saddle}),
\be
{m^z}^C(\beta)= \frac{x_1^C(\beta )-x_4^C(\beta)}{x_1^C(\beta)+x_4^C(\beta) } =p(m^z)^{p-1} \frac{\tanh\left(\beta \sqrt{p^2({m^z}^C)^{2p-2} +\Gamma^2} \right)}{\sqrt{p^2({m^z}^C)^{2p-2} +\Gamma^2} } .
\ee
As the result, the optimal solution is equivalent to the static approximate solution and the static approximation is exact for the infinite-range ferromagnetic $p$-spin model.

%%%%%%%%%%%%%%%%%%%%%%%%%%%%%%%%%%%%%%%%%%%%%%%%%%%%%%%%
%%%%%%%%%%%%%%%%%%%%%%%%%%%%%%%%%%%%%%%%%%%%%%%%%%%%%%%%
%%%%%%%%%%%%%%%%%%%%%%%%%%%%%%%%%%%%%%%%%%%%%%%%%%%%%%%%
{\it General case.}---
Although we have considered the simple system so far, similar analysis is straightforwardly  applicable for generalized mean-field quantum spin systems,
\be 
\hat{H}_G&=&-N \sum_{\mu=1}^k f_\mu \left(\frac{1}{N} \sum_{i=1}^N J_i^\mu  \hat{\sigma}_i^z \right)   -N \sum_{\nu=1}^{l} g_\nu \left( \frac{1}{N} \sum_{i=1}^{N} \Gamma_i^\nu \hat{\sigma}_i^x\right)  \label{general} ,
\ee
where $f_\mu$ and $g_\nu$ are arbitrary functions and $J_i^\mu$ and $\Gamma_i^\nu$ are depending on each site $i$.
Later, we will show that this Hamiltonian contains many mean-field quantum spin systems analyzed in the context of quantum annealing.
The partition function is given by
\be
Z&=&\int \mathcal{D} m_\mu^z \mathcal{D} m_\nu^x \prod_{i=1}^N \left\{  \Tr\left( \prod_{t=1}^M 
e^{ \frac{\beta}{M} \sum_\mu J_i^\mu f_\mu'(m_\mu^z(t))  \hat{\sigma}^z  }
e^{ \frac{\beta}{M} \sum_\nu \Gamma_i^\nu g_\nu'(m_\nu^x(t)) \hat{\sigma}^x  } \right.  \right. 
\no\\ 
&& \left.  \left.   e^{ \frac{\beta}{M}  \left\{ \sum_\mu \left( -f_\mu'(m_\mu^z(t)) \cdot m_\mu^z(t)+ f_\mu(m_\mu^z(t)) \right) + \sum_\nu \left( -g_\nu'(m_\nu^x(t)) \cdot m_\nu^x(t)+  g_\nu(m_\nu^x(t)) \right) \right\} } \right) \right\} .
\ee
From the same discussion as before, we find that, in the thermodynamic limit, the trace of the partition function is evaluated as
\be
&& \lim_{N\rightarrow \infty} \lim_{M\rightarrow \infty} \prod_{i=1}^N \left\{  \Tr\left( \prod_{t=1}^M 
e^{ \frac{\beta}{M} \sum_\mu J_i^\mu f_\mu'(m_\mu^z(t))  \hat{\sigma}^z  }
e^{ \frac{\beta}{M} \sum_\nu \Gamma_i^\nu g_\nu'(m_\nu^x(t)) \hat{\sigma}^x  } \right.  \right. 
\no\\ 
&& \left.  \left.   e^{ \frac{\beta}{M}  \left\{ \sum_\mu \left( -f_\mu'(m_\mu^z(t)) \cdot m_\mu^z(t)+ f_\mu(m_\mu^z(t)) \right) + \sum_\nu \left( -g_\nu'(m_\nu^x(t)) \cdot m_\nu^x(t)+  g_\nu(m_\nu^x(t)) \right) \right\} } \right) \right\} 
\no\\
&=&\lim_{N\rightarrow \infty} \prod_{i=1}^N (x_{4i-3}^\ast(\beta)+x_{4i-3}^\ast(\beta)) ,
\ee
where $x_i^\ast$ is the optimal solution of the following optimal control problem,
\be
\dot{x}_i^\ast(s)&=& \frac{\partial H_{\rm G,op} }{\partial k_i} ( x^\ast(s), k^\ast(s) , {m_\mu^z}^\ast(s), {m_\nu^x}^\ast(s) )  ,
\\
\dot{k}_i^\ast(s)&=& -\frac{\partial H_{\rm G,op} }{\partial x_i} ( x^\ast(s), k^\ast(s) , {m_\mu^z}^\ast(s), {m_\nu^x}^\ast(s))  ,
\\
x_{4j-3}(0) &=& x_{4j}(0) = 1 ,
\\
x_{4j-2}(0) &=& x_{4j-1}(0) = 0 , 
\\
k_{4j-3}(\beta) &=& k_{4j}(\beta) = - \prod_{i=1,  i\neq j }^N (x_{4i-3}(\beta)+ x_{4i}(\beta)) ,
\\
k_{4j-2}(\beta) &=& k_{4j-1}(\beta) = 0 ,
\\
H_{\rm G,op}( x^\ast(s), k^\ast(s) , {m_\mu^z}^\ast(s), {m_\nu^x}^\ast(s) ) 
&=& \min_{m^z}  H_{\rm G,op}( x^\ast(s),  k^\ast(s),{m_\mu^z}^\ast(s), {m_\nu^x}^\ast(s) )  , \label{Gmin}
\ee
where  the classical Hamiltonian $H_{\rm G,op}$ is given by 
\be
H_{\rm op}(x,k,{m_\mu^z},m_\nu^x)
&=&\left\{ \sum_\mu \left( -f_\mu'(m_\mu^z)  m_\mu^z+ f_\mu(m_\mu^z) \right) + \sum_\nu \left( -g_\nu'(m_\nu^x)  m_\nu^x+  g_\nu(m_\nu^x) \right) \right\}  \sum_{i=1}^{4N} x_{i} k_{i} 
\no\\
&& + \sum_\mu \sum_{i=1}^N J_i^\mu f_\mu'(m_\mu^z)  (x_{4i-3} k_{4i-3}-x_{4i-2} k_{4i-2}+x_{4i-1} k_{4i-1}-x_{4i} k_{4i} )
\no\\
&& +\sum_\nu \sum_{i=1}^N  \Gamma_i^\nu g_\nu'(m_\nu^x)   (x_{4i-3} k_{4i-2} + x_{4i-2} k_{4i-3}+x_{4i-1} k_{4i} + x_{4i} k_{4i-1} ) .
\ee
Therefore, finding the exact solution of the partition function is reduced to solving the $4N$-dimensional classical Hamiltonian.
From Eq. (\ref{Gmin}), we find
\be
{m_\mu^z}^\ast(s)&=&\frac{\sum_{i=1}^{N}  J_i^\mu  (x_{4i-3}^\ast k_{4i-3}^\ast-x_{4i-2}^\ast k_{4i-2}^\ast+x_{4i-1}^\ast k_{4i-1}^\ast-x_{4i}^\ast k_{4i}^\ast)  }{\sum_{i=1}^{4N} x_i^\ast k_i^\ast}  ,  \label{Gmin2}
\\
{m_\nu^x}^\ast(s)&=&\frac{\sum_{i=1}^N \Gamma_i^\nu  (x_{4i-3}^\ast k_{4i-2}^\ast + x_{4i-2}^\ast k_{4i-3}^\ast+x_{4i-1}^\ast k_{4i}^\ast + x_{4i}^\ast k_{4i-1}^\ast )  }{\sum_{i=1}^{4N} x_i^\ast k_i^\ast}  ,  \label{Gmin3}
\ee
and Eq. (\ref{Gmin}) is reduced to 
\be
H_{\rm G,op}^\ast&=& \sum_\mu f_\mu \left(\frac{\sum_{i=1}^{N} J_i^\mu  (x_{4i-3} k_{4i-3}-x_{4i-2} k_{4i-2}+x_{4i-1} k_{4i-1}-x_{4i} k_{4i})  }{\sum_{i=1}^{4N} x_i k_i} \right) \sum_{i=1}^{4N} x_i k_i 
\no\\
&&+ \sum_\nu g_\nu\left(\frac{\sum_{i=1}^N \Gamma_i^\nu  (x_{4i-3} k_{4i-2} + x_{4i-2} k_{4i-3}+x_{4i-1} k_{4i} + x_{4i} k_{4i-1} )  }{\sum_{i=1}^{4N} x_i k_i} \right)\sum_{i=1}^{4N} x_i k_i  ,
\no\\ 
\ee
Using the Hamilton equations of $H_{\rm G,op}^\ast$, we find 
\be
\frac{d}{ds}{m_\mu^z}^\ast(s)&=&-2 \frac{ \sum_\nu \sum_{i=1}^{N}g_\nu' \Gamma_j^\nu (x_{4i-3}^\ast k_{4i-2}^\ast-x_{4i-2}^\ast k_{4i-3}^\ast+x_{4i-1}^\ast k_{4i}^\ast-x_{4i}^\ast k_{4i-1}^\ast)  }{\sum_{i=1}^{4N} x_i^\ast k_i^\ast} , 
\\
\frac{d}{ds} {m_\nu^x}^\ast(s)&=& 2 \frac{ \sum_\mu \sum_{i=1}^{N}f_\mu' J_j^\mu  (x_{4i-3}^\ast k_{4i-2}^\ast-x_{4i-2}^\ast k_{4i-3}^\ast+x_{4i-1}^\ast k_{4i}^\ast-x_{4i}^\ast k_{4i-1}^\ast)  }{\sum_{i=1}^{4N} x_i^\ast k_i^\ast} . 
\ee
Then, from the same technique as before, we can show that the optimal solution is equal to the static approximate solution and Eqs. (\ref{Gmin2}) and (\ref{Gmin3}) reproduce the saddle point equations of the static approximation,
\be
{m_\mu^z} &=& \frac{1}{N}\sum_{i=1}^{N}  J_i^\mu \frac{ x_{4i-3}(\beta) -x_{4i}(\beta)  }{ x_{4i-3}(\beta)+x_{4i}(\beta)} 
\no\\
&=& \frac{1}{N}\sum_{i=1}^{N}  J_i^\mu  \frac{ \left( \sum_\mu J_i^\mu f_\mu'(m_\mu^z)\right)  \tanh \left(\beta \sqrt{  (\sum_\mu J_i^\mu f_\mu'(m_\mu^z) )^2   + (\sum_\nu \Gamma_i^\nu g_\nu'(m_\nu^x)  )^2  } \right )}{\sqrt{  (\sum_\mu J_i^\mu f_\mu'(m_\mu^z) )^2   + (\sum_\nu \Gamma_i^\nu g_\nu'(m_\nu^x)  )^2  } } , 
\\
{m_\nu^x} &=& \frac{1}{N}\sum_{i=1}^{N}  \Gamma_i^\nu \frac{x_{4i-2}(\beta)+x_{4i-1}(\beta)   }{ x_{4i-3}(\beta)+x_{4i}(\beta)} 
\no\\
&=& \frac{1}{N}\sum_{i=1}^{N}  \Gamma_i^\nu  \frac{ \left(\sum_\nu \Gamma_i^\nu g_\nu'(m_\nu^x) \right)  \tanh \left(\beta \sqrt{  (\sum_\mu J_i^\mu f_\mu'(m_\mu^z) )^2   + (\sum_\nu \Gamma_i^\nu g_\nu'(m_\nu^x)  )^2  } \right )}{\sqrt{  (\sum_\mu J_i^\mu f_\mu'(m_\mu^z) )^2   + (\sum_\nu \Gamma_i^\nu g_\nu'(m_\nu^x)  )^2  } }  .
\ee
Therefore, the static approximation is exact for generalized mean-field quantum spin systems (\ref{general}) in general.

%%%%%%%%%%%%%%%%%%%%%%%%%%%%%%%%%%%%%%%%%%%%%%%%%%%%%%%%%
%%%%%%%%%%%%%%%%%%%%%%%%%%%%%%%%%%%%%%%%%%%%%%%%%%%%%%%%
{\it Application to quantum annealing.}---
We apply our result to previous studies of quantum annealing, where mean-field quantum spin systems are often used to evaluate the performance of quantum annealing \cite{JKKMP,SN,SN2,SYYN,MNL}, and our result guarantees their analysis in general.
For example, we consider  the Hopfield model with finite-number patterns,
\be 
\hat{H}&=&-N \sum_{\mu=1}^k  \left(\frac{1}{N} \sum_{i=1}^N J_i^\mu  \hat{\sigma}_i^z \right)^p  -\Gamma_1  \sum_{i=1}^{N} \hat{\sigma}_i^x  +\Gamma_2 N \left( \frac{1}{N} \sum_{i=1}^{N} \hat{\sigma}_i^x\right)^2  \label{hop}
\ee
where $\Gamma_1, \Gamma_2\ge0$, $p$ is an integer denoting the degree of interactions, and $k$ is an integer representing the finite-number embedded pattern and $J_i^\mu$ takes $\pm1$ at random.
The antiferromagnetic multiple-$X$ term is called the non-stoquastic interaction and has been attracting a lot of attention in the field of quantum annealing in recent years \cite{BDOT,NT}.
In the process of quantum annealing, recent studies \cite{SN,SN2} show that, although this system undergoes a first-order phase transition in the absence of the antiferromagnetic multiple-$X$ term,  there is a path through a second-order phase transition avoiding a first-order phase transition when  the antiferromagnetic multiple-$X$ term is applied.
This means that the antiferromagnetic multiple-$X$ term improves exponentially the efficiency of quantum annealing compared with the case only by the transverse-field.
Although the above result is based on the static approximation, the Hamiltonian (\ref{hop}) is included in our Hamiltonian (\ref{general}) and, thus, the result of exponential speed up by the non-stoquastic interaction is exact.

Next, we consider the infinite-range ferromagnetic $p$-spin model with longitudinal random field,
\be
\hat{H}= -  N \left(\frac{1}{N^{}} \sum_i \hat{\sigma}_i^z \right)^p -  \sum_i h_i \hat{\sigma}_i^z  - \sum_i \Gamma_i \hat{\sigma}_i^x ,
\label{inhom}
\ee
where  $h_i$ follows the Gaussian distribution with an average of $0$ or the binary distribution $h_i=\pm h_0$.
Based on the static approximation, a recent study \cite{SYYN} shows that the inhomogeneous transverse field $\Gamma_i$ can avoid a phase transition in the process of quantum annealing, although the homogeneous  transverse field can not avoid a first-order phase transition.
This means that the inhomogeneous transverse field accelerates  the computation time of quantum annealing exponentially.
Our model (\ref{general}) includes the above system (\ref{inhom}) and certifies the analysis based on the static approximation.

%%%%%%%%%%%%%%%%%%%%%%%%%%%%%%%%%%%%%%%%%%%%%%%%%%%%%%%%%
%%%%%%%%%%%%%%%%%%%%%%%%%%%%%%%%%%%%%%%%%%%%%%%%%%%%%%%%
{\it Conclusions.}---
We have obtained the exact solution of the partition function for the particular class of mean-field quantum spin systems including randomness and showed that the static approximation is exact for their models in general.
Although the imaginary-time dependence of the partition function of mean-field quantum spin systems had not been analyzed exactly,  we gave a method to solve this problem exactly.
Our result demonstrates that the method of the optimal control problem is a powerful approach to analysis of statistical mechanics.

As an application to quantum annealing, we verified exactness of the result of exponential speed up of quantum annealing by the non-stoquastic interaction or the inhomogeneous transverse field, which have recently attracted a lot of interest.
Our analysis is an effective  approach to quantum annealing by statistical mechanics.

It is an interesting and important future problem to obtain the exact solution for the case where the static approximation is not exact.
Previous studies \cite{YI,WW,GL,Goldschmidt,UBK,MH,ONS,Takahashi} show that the static approximation gives non-physical solution for the $p$-spin-interacting spin glass model in the transverse field which has the spin glass phase.
Our method may be the first step to exactly analyze such system where the static approximation is broken.

%%%%%%%%%%%%%%%%%%%%%%%%%%%%%%%%%%%%%%%%%%%%%%%%%%%%%%%%%%%%%%%%%%%%%%%%%%%%
{\it Acknowledgments.}---
The authors are deeply grateful to Keisuke Fujii, Shunji Matsuura, Hidetoshi Nishimori, Kohji Nishimura, Kentaro Ohki and Kabuki Takada for useful discussions.
M. Okuyama was supported by JSPS KAKENHI Grant No. 17J10198.
M. Ohzeki was supported by ImPACT Program of Council for Science, Technology and Innovation (Cabinet Office, Government of Japan) and JSPS KAKENHI No. 15H03699 No. 16K13849, No. 16H04382.

%%%%%%%%%%%%%%%%%%%%%%%%%%%%%%%%%%%%%%%%%%%%%%%%%%%%%%%%%%%%%%%%%%%%%%%%%%%%
%%%%%%%%%%%%%%%%%%%%%%%%%%%%%%%%%%%%%%%%%%%%%%%%%%%%%%%%%%%%%%%%%%%%%%%%%%%%

\onecolumngrid

\def\theequation{S\arabic{equation}}
\makeatletter
\@addtoreset{equation}{section}
\makeatother

\setcounter{equation}{0}

%\preprint
\newpage
\section{Supplemental material for \\
``An exact solution of the partition function for mean-field quantum spin systems without the static approximation''}

\twocolumngrid
%%%%%%%%%%%%%%%%%%%%%%%%%%%%%%%%%%%%%%%%%%%%%%%%%%%%%%%%%%%%%%%%%%%%%%%%%%%
\section{Optimal control problem}
In this section, we provide some knowledge about the optimal control problem.

For $m$ differential equations describing the motion of $m$ dimensional vector $x(s)$,
\be
\dot{x} (s) = f(x(s), u(s)) ,  \  x(0)=x_0,
\ee
let us consider the problem of finding the control input $u(s)\in U(0,T)$ that minimizes the cost function,
\be
J= L_f(x(T)) + \int_0^T ds L(x(s), u(s)) ,
\no\\
\ee
where the initial state $x_ 0$ and the final time $T$ are known and the final state $x(T)$ is arbitrary.
Here, using the auxiliary variable $k$ which is the $m$-dimensional vector, we define the following function,
\be
H_{\rm op} =k^{\mathrm{T}} f(x,u)  -L (x,u),
\ee
then, the following result holds.

A necessary condition for the control input $u^\ast(t)$ and the corresponding trajectory $x^\ast(t)$ to be optimal is that there exists the function $k^\ast(t)$ that simultaneously satisfies the following three conditions.

(a) $x^\ast(t)$ and $k^\ast(t)$ are solutions to the following ordinary differential equations,
\be
\dot{x}^\ast(s)&=& \frac{\partial H_{\rm op} }{\partial k} ( x^\ast(s), u^\ast(s), k^\ast(s) )  ,
\\
\dot{k}^\ast(s)&=& -\frac{\partial H_{\rm op} }{\partial x} ( x^\ast(s), u^\ast(s), k^\ast(s) ) .
\ee
(b) $k^\ast(t)$ satisfies the following boundary condition,
\be
k^\ast(T)=  \frac{\partial L_f (x, u) }{\partial x}  |_{x=x^\ast (T)} .
\ee
(c) For any time $ s \in [0, T] $,
\be
H_{\rm op}( x^\ast(s), u^\ast(s), k^\ast(s) ) = \min_{u\in U}  H_{\rm op}( x^\ast(s), u, k^\ast(s) )  .
\no
\ee

\end{document}